\documentstyle[epsf,aps]{revtex}

\begin{document}

\twocolumn[\hsize\textwidth\columnwidth\hsize\csname
@twocolumnfalse\endcsname

\title{Probing dark energy with gamma-ray bursts}
\author{Keitaro Takahashi$^{1}$, Masamune Oguri$^{1}$, 
Kei Kotake$^{1}$ and Hiroshi Ohno$^{2,3}$ }

\address {$^{1}$Department of Physics, University of Tokyo, 
7-3-1 Hongo, Bunkyo, Tokyo 113-0033, Japan}

\address{$^{2}$Division of Theoretical Astrophysics,
National Astronomical Observatory,
2-21-1 Osawa, Mitaka, Tokyo 181-8588, Japan}

\address{$^{3}$Research Center for the Early Universe (RESCEU),
University of Tokyo, 7-3-1 Hongo, Bunkyo, Tokyo 113-0033, Japan} 

\date{\today}

\maketitle

\begin{abstract}

We propose a new method to use gamma-ray bursts (GRBs) as an alternative
probe of the dark energy. By calibrating luminosity-variability and
luminosity-lag time relations at low redshift where distance-redshift
relations have been already determined from type Ia supernova (SNIa), GRBs at
high redshift can be used as a distance indicator. We investigate the
potential impact of future GRB data on determining the current matter
density $\Omega_{m}$ and the dark energy equation of state $w$ which is
assumed to be constant. We show that a combined analysis of a data set
expected from the {\it Swift} and the current SNIa data results in
 excellent determination of both  $\Omega_{m}$ and $w$. 
\end{abstract}
\vskip2pc]


\vskip1cm


Accelerating universe, which is strongly suggested by the distance
measurements of type Ia supernova (SNIa) \cite{SCP1,SCP2,Riess1998}, 
is one of the most interesting and challenging problems in modern
cosmology. Recent observation of the cosmic microwave background (CMB)
anisotropy \cite{MAP} revealed that the universe is almost flat and
confirmed the need for the energy component which accelerates our
universe. 

The cosmological constant, or ``dark energy'' in general, is often
introduced to account for this cosmic acceleration. Dark energy is the
unknown energy component which has effective negative pressure $p = w
\rho$ ($w < 0$). The case with $w = -1$ corresponds to the cosmological
constant. Since the value of $w$, which is a function of redshift $z$ in
general, is closely related to the nature of dark energy, it is of great
importance to determine the value of $w$ observationally. Until now, a
number of methods are proposed to probe the dark energy equation of state
\cite{Turner1997,Lui1999,Haiman2001,Weller2001,Hu2002,Matsubara2002,Munshi2003,Blake2003,Melchiorri2002}.
However, currently $w$ has not been determined quite well even if one
assumes that $w$ is constant \cite{Hannestad2002,Spergel2003}. In
particular, the current distance measurements of SNIa poorly determines
the value of $w$ because of the degeneracy between $w$ and the
dark-energy density \cite{SCP1,SCP2,Riess1998}. Moreover, the use of CMB
anisotropy, which is known to be one of the most powerful and clean
probe of cosmological parameters, introduces an additional degeneracy
between the Hubble constant and dark-energy parameters
\cite{Spergel2003}. That is one of the reason why future SNIa search,
such as  the proposed {\it SNAP} satellite \cite{SNAP}, has been
vigorously studied as a probe of the dark energy
\cite{HutererTurner1999,WellerAlbrecht,ChibaNakamura2000}. In this {\it
Letter}, we propose another method using gamma-ray bursts (GRBs) as
a distance indicator.  

Recent studies have suggested the possibilities that GRBs can be
standard candles through the relations between absolute
``isotropic-equivalent'' luminosity ($L$) and other observed quantities.
Two of them, which we use in this {\it Letter}, are variability ($V$) 
\cite{FenimoreRamirezRuiz2000,Reichart2001} and spectral lag 
($\tau_{\rm lag}$) \cite{NorrisMaraniBonnell2000,Salmonson2001}.  
The variability is a specific measure of the ``spikiness'' of the light
curve. The spectral lag is the time between peaks as recorded at high
and low photon energies. These two quantities are related to the
luminosity as $L \propto V^{\alpha_{V}}$ and $L \propto \tau_{\rm
lag}^{\alpha_{\rm lag}}$, respectively. Furthermore, a $V/\tau_{\rm
lag}$ relation, which must exist if the $L/V$ and $L/\tau_{\rm lag}$
relations are true, has been confirmed with an independent sample of 112
BATSE bursts \cite{SchaeferDengBand2001}. There are also some
theoretical explanations for these relations
\cite{IokaNakamura2001,Plaga2000}. Once these relations are established
by observations, then one can use these relations to derive the absolute
luminosity of GRBs and to estimate the luminosity distance to the GRBs
\cite{Schaefer2003}. 

Up to now, only $\sim 10$ GRBs have the enough information ($z, V,
\tau_{\rm lag}$) to establish the above two relations. However the {\it
Swift} satellite \cite{Swift}, planned for launch in 2003, is expected
 to detect $\sim 1000$ GRBs, about half of which 
 will be observed  with known redshifts,
 during 3 years of observations. This huge number of GRBs allows us to
probe the expansion history of the universe more deeply than SNIa
\cite{Schaefer2003}. In this {\it Letter}, we simulate the data which is
expected to be obtained by the {\it Swift} and investigate the ability
to probe the dark energy independently with other methods such as CMB
anisotropy. Throughout the {\it Letter}, we assume a flat universe
$\Omega_{m}+\Omega_{X}=1$, where $\Omega_{m}$ is the present matter
density and $\Omega_{X}$ is the present dark-energy density.

Here we must be careful to avoid circular logic. The original studies of 
the $L/V$ and $L/\tau_{\rm lag}$ relations assumed a particular set of 
cosmological parameters (the Hubble constant $H_{0}$, $w$, $\Omega_{m}$
and the flatness of the universe) to derive the relations. Without any
assumptions about the cosmological parameters, we cannot determine $L/V$
and $L/\tau_{\rm lag}$ relations which are used to obtain the absolute
luminosity of GRBs. Thus we assume that a form of the luminosity
distance as a function of the redshift at $z < 1$ has already been well
determined by the current SNIa observations. This is approximately that
of the $\Lambda$-dominated universe ($w = -1, \Omega_{X} = 0.7$ and
$\Omega_{m} = 0.3$). The errors associated with the form of the
luminosity distance at $z < 1$ is reflected in the systematic errors in
the derived absolute luminosities of the low-$z$ ($z < 1$) GRBs. We
calibrate these two relations by using only low-$z$ GRBs, and obtain the
absolute luminosities of high-$z$ ($z > 1$) GRBs using these relations.
After that, the procedure is the same as that using SNIa: fitting
parameters such as $w$ and $\Omega_{m}$ so as to reproduce the magnitude
as a function of the redshift. The apparent magnitude $m$ and redshift
$z$ are related by the luminosity distance as: 
\begin{equation}
m = M + 25 + 5 \log{d_{L}(z)},
\end{equation}
where $M$ is an absolute magnitude and 
\begin{eqnarray}
d_{L}(z) & \equiv & \frac{(1+z) c}{H_{0}} \int^{z}_{0} dz' \nonumber \\ 
& & \times\left[
\Omega_{m} (1+z')^{3} + (1 - \Omega_{m}) (1+z')^{3+3w} 
\right]^{-\frac{1}{2}} 
\end{eqnarray}
is the luminosity distance in units of Mpc.

\begin{figure}[h]
\epsfxsize=9cm
\centerline{\epsfbox{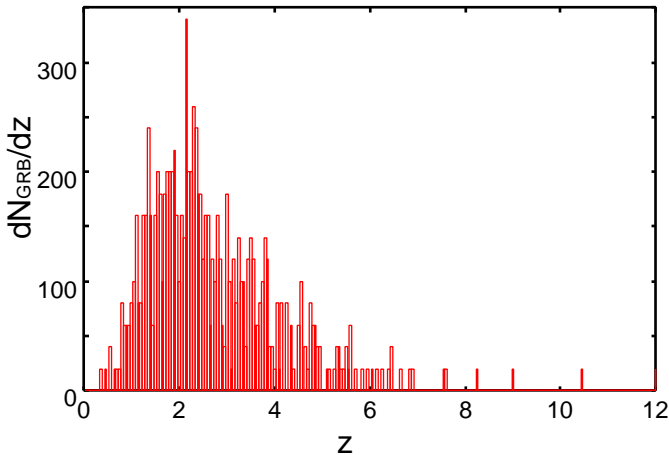}}
\caption{Redshift distribution of the simulated GRBs. The flux limit of
 the {\it Swift} is applied.}
\label{fig:z_distribution}
\end{figure}
 
For concreteness, we simulate a future data set similar to one expected
from the {\it Swift}. The sample comprises 500 GRBs, which is consistent
with the expected number of GRBs with measured redshifts. Their
redshifts and luminosities are distributed according to the GRB rate
history and the luminosity function based on the model of the star
formation rate 2 (SF2) in  \cite{PorcianiMadau2001}, which is roughly
constant at $z \gtrsim 2$.  This model is consistent with the current
observations of UV luminosity density of the galaxy population. 
The energy spectrum of the GRB is assumed to be a single power-law with index
$-2.25$. The observed luminosities are calculated assuming a background
cosmology with $H_{0} = 65 {\rm km\,s^{-1}Mpc^{-1}}, \Omega_{m} = 0.3,
\Omega_{X} = 0.7$ and $w = -1$. The flux limit of the {\it Swift}, 
$f > 0.04 \; {\rm photons\,cm^{-2}s^{-1}}$, is applied to check whether a
produced GRB will be observed or not. The errors in redshifts
are tentatively neglected because of the lack of information on the
redshift measurement by {\it Swift}, while we will show later that  
inclusion of the errors in redshifts does not change our result so much.
We assume the log-normal distribution of the luminosity errors of 0.02
in logarithmic units of base 10. Other errors below are also assumed to
be the log-normal distribution. Fig. \ref{fig:z_distribution} shows the
redshift distribution of the simulated GRBs. The average redshift of
these GRBs is about 3. Although very high-$z$ GRBs might be subject to
some possible biases, they are practically not influential on the
analyses below, because the number of such high-$z$ GRBs is not so
large. 

\begin{figure}[h]
\epsfxsize=9cm
\centerline{\epsfbox{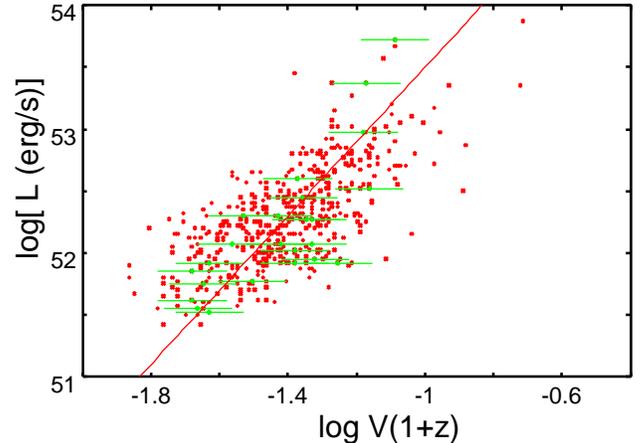}}
\caption{Absolute luminosity-variability relation for the simulated
data. Points with error bars show low-$z$ ($z < 1$) GRBs, which
are used to derive $L/V$ relation. Points without errorbars
show high-$z$ ($z > 1$) GRBs. Note that the high-$z$ GRBs also have
the same errors as low-$z$ GRBs. Solid line shows the assumed
$L/V$ relation (\ref{eq:L-V}).}
\label{fig:L-V}
\end{figure}

\begin{figure}[h]
\epsfxsize=9cm
\centerline{\epsfbox{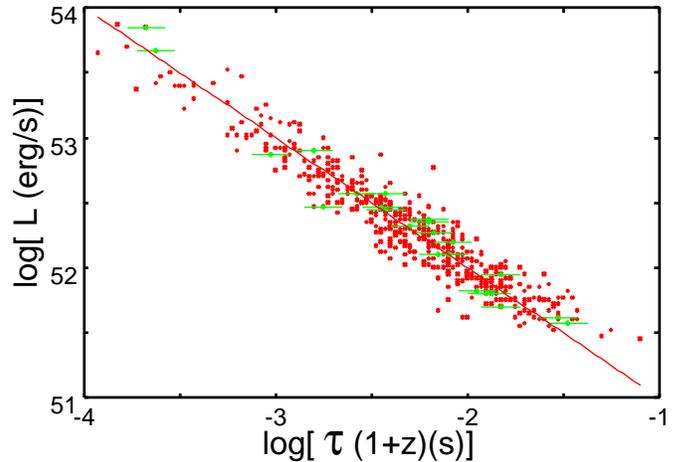}}
\caption{Same as Fig. \ref{fig:L-V}, but the absolute luminosity-spectral
 lag relation is plotted. Solid line shows the assumed
$L/\tau_{\rm lag}$ relation (\ref{eq:L-t}).}
\label{fig:L-t}
\end{figure}

Next the observed variability and spectral lag are given to each GRB 
according to the following $L/V$ and $L/\tau_{\rm lag}$ relations,
\begin{eqnarray}
L & = & 3 \times 10^{56} \left( V(1+z) \right)^{3} {\rm erg / s}, 
\label{eq:L-V} \\
L & = & 10^{50} \left( \frac{\tau_{\rm lag} (1+z)}{1 {\rm s}} \right)^{-1} 
{\rm erg / s} \label{eq:L-t},
\end{eqnarray}
with the intrinsic scatter as well as the measurement uncertainties.
For both the $L/V$ and $L/\tau_{\rm lag}$ relations, we assume both the
intrinsic scatter and measurement error to be 0.1 in logarithmic units.
The indices and proportional constants of these relations are different
among various studies. The values we use are typical of them
\cite{FenimoreRamirezRuiz2000,Reichart2001,NorrisMaraniBonnell2000,Salmonson2001}.
The measurement errors are almost the same as those of the current 
observations. The simulated data are plotted in Fig. \ref{fig:L-V} and
\ref{fig:L-t} with the assumed relations (\ref{eq:L-V}) and
(\ref{eq:L-t}), respectively. Although the $L/\tau_{\rm lag}$ relation
might steepen to  $L \propto \tau_{\rm lag}^{-3}$ at $\tau_{\rm lag}
\gtrsim 0.1 {\rm s}$ \cite{Salmonson2001}, our assumption of single
power-law for the $L/\tau_{\rm lag}$ relation is enough because GRB
events with such large $\tau_{\rm lag}$ will be very rare for the
luminosity function adopted here. In fact the indices that we adopted
need not be the true values since they are to be determined
from the {\it Swift} observation. Thus what we require are that
there are some one-to-one relations between the observable quantities 
($V$ and $\tau_{\rm lag}$) and the absolute luminosity $L$ and that
the relations are independent on z for $0 < z \lesssim 3$. 
Although evolution of the relations is not likely as is stated 
in \cite{Schaefer2003}, it will be tested if the luminosity distance 
up to $z \lesssim 1.7$ is obtained from the {\it SNAP} observation.

\begin{figure}[h]
\epsfxsize=9cm
\centerline{\epsfbox{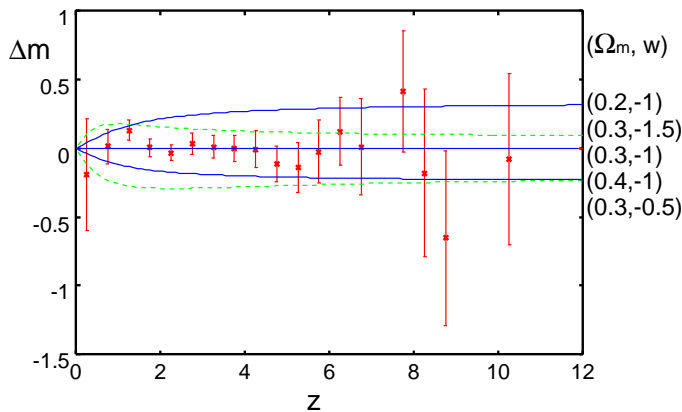}}
\caption{Magnitude difference $\Delta m$ between several cosmological 
models and the fiducial model ($\Omega_{m}=0.3$, $w=-1$), along with the
 simulated binned data. Three solid curves are for $(\Omega_{m},w) = (0.2,-1)$,
 $(0.3,-1)$ and $(0.4,-1)$, and two dashed curves are for 
$(\Omega_{m},w) = (0.3,-1.5)$ and $(0.3,-0.5)$.} 
\label{fig:hubble_diagram}
\end{figure}

Then the absolute luminosities of the low-$z$ GRBs are calculated
by the luminosity distance-redshift relation which has been determined
by the current SNIa observations. Its uncertainties lead to the
uncertainties of the absolute luminosities of $\sim 0.2$ in logarithmic
units. Then the $L/V$ and $L/\tau_{\rm lag}$ relations are derived by the
low-$z$ GRBs. As a result, we recover the the $L/V$ and $L/\tau_{\rm
lag}$ relations of the form $L\propto V^{2.13}$ and $L\propto \tau_{\rm
lag}^{-0.99}$ from the simulated low-$z$ GRBs. Two independent absolute
luminosities of high-$z$ GRBs are obtained by these relations, and are
combined as weighted averages to produce combined absolute luminosities.
Fig. \ref{fig:hubble_diagram} shows the binned magnitude difference
$\Delta m$ between several representative cosmological models and the
fiducial model ($\Omega_{m}=0.3$, $w=-1$), along with the simulated data. 

\begin{figure}[h]
\epsfxsize=9cm
\centerline{\epsfbox{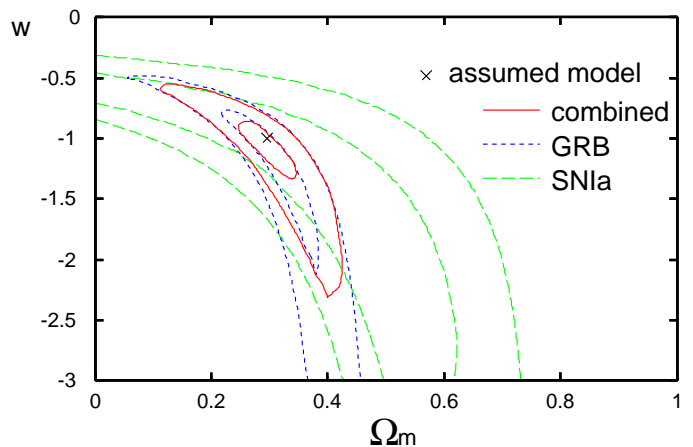}}
\caption{Best fit $1 \sigma$ and $3 \sigma$ confidence regions in the 
$\Omega_{m}-w$ plane by the simulated GRB data (dotted), the current SNIa 
data (dashed) and the combined analysis (solid). The cross shows the
assumed model $(\Omega_{m},w)=(0.3,-1)$. A flat universe is assumed.}
\label{fig:contour}
\end{figure}

Finally we perform $\chi^2$ constraints to see to what extent we can
constrain dark energy properties from the observation of GRBs. Here we
assume the dark energy to have a time-independent equation of state $w =
p/\rho = {\rm const}$. We do not put a prior  $w \geq -1$ both because
it may bias our final result \cite{Hannestad2002} and because several
theoretical models which predict $w<-1$ do exist
\cite{Caldwell2002,Chiba1999}. Further a flat universe with $H_{0} = 65
{\rm km\,s^{-1}Mpc^{-1}}$ is assumed. We calculate $\chi^{2}$ as
\begin{equation}
\chi^{2} \equiv \sum_{i}^{N_{\rm GRB}}
\frac{\left[m_i - M_i - 25 - 5 \log{d_{L}(z_{i},\Omega_{m},w)}\right]^2}{\delta m_i^2},
\end{equation}
where $N_{\rm GRB}$ is the number of GRBs, $m_i$ is the observed apparent
magnitude, $M_i$ is the absolute magnitude estimated from $L/V$ and
$L/\tau_{\rm lag}$ relations, and $\delta m_i$ is the error of the
magnitude. SNIa data fit and the combined fit of GRB and SNIa data are
also performed. For the SNIa data, we use the data of the SCP (primary
data set C) \cite{SCP2}, the High-Z Search Team \cite{Riess1998} and
Cal\'{a}n-Tololo survey \cite{Hamuy1993} for the SNIa analysis. Fig.
\ref{fig:contour} shows the best-fit $1 \sigma$ and $3 \sigma$
confidence regions in the $\Omega_{m}-w$ plane by the simulated GRB
data, the current SNIa data and the combined analysis. Since average
redshifts are significantly different between SNIa ($\sim 0.5$) and GRB
($\sim 3$), the combined fit results in the excellent determination of
both $\Omega_{m}$ and $w$. Our simulation demonstrates that $\Omega_{m}$
and $w$ can be determined within $\sim 20$\% accuracy from the
combined analysis of only two types of observation;, SNIa and GRBs. Note
that constraints from both SNIa and GRBs are quite insensitive to other
(cosmological) parameters, such as the Hubble constant, the spectral
index of primordial power spectrum, and the normalization of rms mass
fluctuations, which are also very important in using other methods,
e.g., CMB anisotropy or cluster abundance.

\begin{figure}[h]
\epsfxsize=9cm
\centerline{\epsfbox{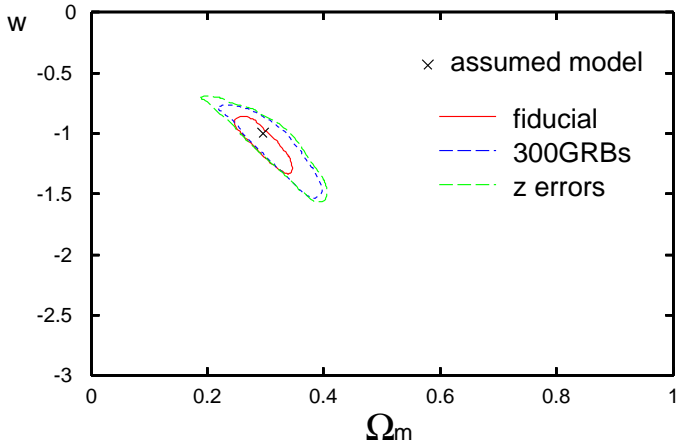}}
\caption{Best fit $1 \sigma$ confidence regions of the combined analysis
 for two additional pessimistic simulations as well as the fiducial
 simulation. Solid, dotted and dashed curves correspond to the result of
 the fiducial simulation, one with 300 GRBs, and one with redshift error
 of 0.2 in logarithmic units, respectively. The cross shows the assumed
 model $(\Omega_{m},w)=(0.3,-1)$.} 
\label{fig:one-sigma}
\end{figure}

Besides the fiducial simulation, we perform two additional pessimistic
simulations: one with 300 GRBs instead of fiducial 500 GRBs, and another
with redshift error of 0.2 in logarithmic units which is neglected in
Fig. \ref{fig:contour}. The best fit $1 \sigma$ confidence regions for
these two pessimistic simulations as well as the fiducial simulation are  
shown in Fig. \ref{fig:one-sigma}. Although the two pessimistic
simulations result in larger $1 \sigma$ regions, $\Omega_{m}$ and $w$
are still determined well. Therefore, we conclude that search for
distant GRBs can be a promising way to probe the dark energy density and dark
energy equation of state separately.  

In summary, we have investigated a potential impact of future GRB
searches on probing the cosmological parameters such as the matter
density fraction of the universe $\Omega_{m}$ and the equation of state 
of dark energy $w$. Our result is that a combined analysis of
a data set expected from the {\it Swift} and the current SNIa data
results in the excellent determination of these parameters. 

K.T.'s work is supported by Grant-in-Aid for JSPS Fellows.
We thank B. Schaefer for fruitful discussions.

\end{document}